# A PubMed-Scale Dataset of Structured Biomedical Abstracts

Chia-Hsuan Chang[1]*, Haerin Song[2]*, Brian Ondov[1], Hua Xu[1]

[1]Department of Biomedical Informatics & Data Science, School of Medicine, Yale University, New Haven, CT 06510, USA

[2]Interdisciplinary Program in Artificial Intelligence, Seoul National University, 1, Gwanak-ro, 08826, Seoul, Republic of Korea

*These authors contributed equally

Corresponding Author: Hua Xu (hua.xu@yale.edu)

**Abstract:** Structured abstracts are important for biomedical literature processing, by facilitating information retrieval, text mining, and knowledge synthesis. However, a vast portion of abstracts indexed in PubMed remain unstructured, presenting a significant bottleneck for downstream text-processing workflows and applications. To resolve this limitation, we introduce *Structured PubMed*, a comprehensive corpus of section-labeled biomedical abstracts compiled from the complete PubMed database, encompassing over 23.2 million research-article records. The corpus is divided into two distinct subsets: a collection of 5.9 million author-structured abstracts parsed from official XML files, and an automatically labeled collection of 17.2 million originally unstructured abstracts structured via a verbatim-extraction Large Language Model pipeline. Every record is harmonized under a unified five-section schema and mapped to its original PubMed identifier, publication type, and publication date. This dataset can be utilized to train sentence-classification models, benchmark text-segmentation architectures, and perform large-scale, section-specific information extraction at an unprecedented PubMed-wide scale.

## Background & Summary

Structured abstracts, in which the text is partitioned into labeled sections such as Background, Objective, Methods, Results, and Conclusions, facilitate information retrieval, information extraction, and knowledge synthesis from the biomedical literature[1,2]. Their section structure has been used to build sentence-classification datasets for medical abstracts[3], to detect the novelty of biomedical findings from conclusion sections[4], and to support large-scale information extraction and evidence synthesis[5], among other applications. The US National Library of Medicine (NLM) encourages this format, which is based on the IMRAD convention[6] and earlier proposals for more informative medical abstracts[7] and is recommended by the ICMJE[8]. NLM maintains a mapping of heading variants onto a small set of standardized categories[9]. Despite this, a substantial portion of abstracts indexed in PubMed are distributed without explicit section labels[10,11]. This creates a practical bottleneck for section-aware retrieval, evidence extraction, and large-scale biomedical NLP, because methods developed on structured abstracts[4] cannot be applied directly to the much larger body of unstructured abstracts without additional labeling.

Recovering this structure has been studied primarily as a sentence classification problem[1,3,10,11]. Hu et al. proposed a few-shot prompt-learning approach[1] that classifies sentences in randomized controlled trial (RCT) and observational study (OS) abstracts into the standardized sections, reporting overall F1 scores of 0.9508 and 0.9401 on the PubMed 200k and 20k RCT benchmarks and improved generalization across study types while using less annotated data. Such methods rely on benchmark corpora derived from already-structured abstracts, and the resulting labeled resources have largely been confined to particular study types and to the tens- or hundreds-of-thousands scale.

A separate line of work has focused on releasing large collections of structured abstracts for downstream modeling tasks. MedConclusion[2] assembles 5.7M PubMed structured abstracts to study biomedical conclusion generation, pairing the non-conclusion sections of each abstract with its author-written conclusion as naturally occurring supervision. That resource is curated specifically for the conclusion-inference task and is drawn from abstracts that were already structured by their authors.

In this study, we describe a corpus of section-labeled biomedical abstracts, *Structured PubMed*, assembled from the complete PubMed baseline and update files, comprising 23.3M research-article records in total. The corpus is organized into two complementary subsets. The first contains 5,993,846 records whose section labels are author-provided, parsed directly from the NLM PubMed XML files, using the NLM-assigned section category where available. The second contains 17,287,922 records whose abstracts were published without section labels and to which section structure is labeled using a large language model (LLM) under a verbatim-extraction protocol (described in Methods). The verbatim-extraction design constrains the model to reproduce only text present in the source abstract, and corpus-level quality is reported in the Technical Validation section. By assigning section structure to abstracts that were not originally structured, the corpus extends section labels to a larger portion of PubMed than resources restricted to natively structured abstracts. Both subsets are released in a common schema keyed on PubMed identifiers, alongside publication type and date. To the best of our knowledge, this is the largest corpus of section-labeled biomedical abstracts released to date, approximately 4.09 times the size of MedConclusion[11].

To summarize, *Structured PubMed* is intended to support reuse across several settings. The author-provided subset offers human-curated structured abstracts for training and evaluating abstract-structuring models, including the sentence-classification approaches noted above. The LLM-labeled subset extends section structure to abstracts that previously lacked it. Together, the two subsets enable users to compare model behavior across natively structured and retrospectively structured abstracts and to perform section-specific retrieval, summarization, and evidence extraction over records that prior labeled resources did not cover, complementing resources such as MedConclusion. To ensure the traceability to source PubMed records and integrate with other PubMed-derived datasets (e.g., iCite[12] and PubTator 3.0[13]), *Structured PubMed* retains PubMed identifier, publication type, and publication date for every structured abstract.

# Methods

## Input Data

The primary source for this dataset is the complete NLM PubMed XML distribution, accessed via the PubMed FTP server[14] on May 3, 2026. The download comprises two components following NLM's recommended loading procedure: (1) the annual baseline snapshot, consisting of 1,334 compressed XML files, and (2) the update files, consisting of 98 additional compressed XML files representing new, revised, and deleted citations made available between the baseline release and the download date. Together, these files contain 40,451,610 citation records.

## PubMed Data Collection & Preprocessing

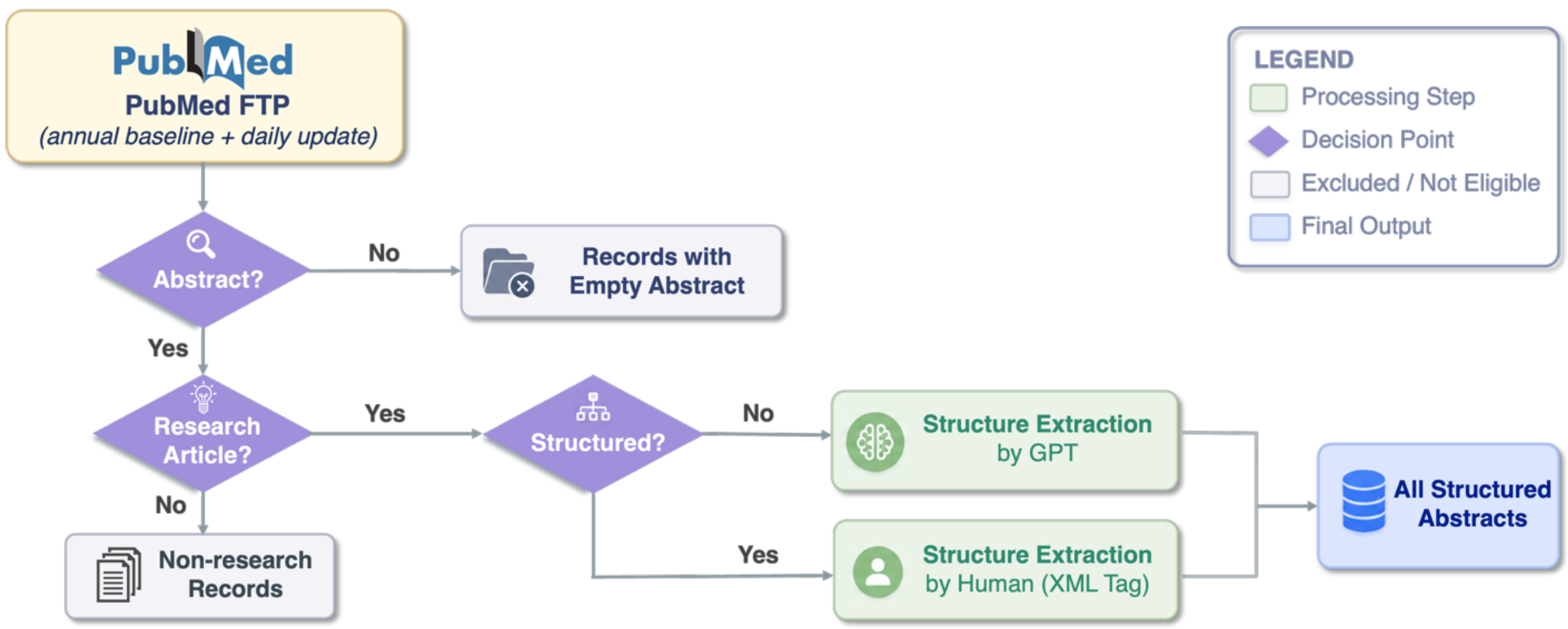

*Figure 1. The workflow of data collection and preprocessing*

We follow the workflow described in Figure 1 to process the PubMed collection. First, records with empty abstract are excluded, reducing the size of collection to 28,705,205 records. Second, review articles and records tagged with any of the non-research article type defined in Table 1 are removed. This step removes 5,344,252 records and retains 23,360,953 research article records in the collection.

*Table 1. Selected non-research NLM publication types*

| Publication Types |
| --- |
| Letter, Case Report, Technical Report, Meta-Analysis, Video-Audio Media, Dataset, Personal Narrative, Preprint, Biography, Historical Article, Portrait, Legal Case, Conference/Congress proceedings, Interview, Address, Bibliography, Legislation, Lecture, Guideline, Directory, News, Webcast, Festschrift, Dictionary, Interactive Tutorial, Patient Education Handout, Editorial, Expression of Concern, Introductory Journal Article, Published Erratum, Corrected and Republished Article, Duplicate Publication, Retracted Articles, and Comment |

Lastly, for every research article record, we determine whether a structured abstract is present by parsing each record's XML attributes. An abstract is classified as structured if at least one non-empty *NLMCategory* or *Label* attribute in *AbstractText* element is identified. When both attributes are present on the same element, *NLMCategory* is preferred. NLM define *NLMCategory*[9] based on IMRAD format, and the available sections include BACKGROUND, OBJECTIVE, METHODS, RESULTS, and CONCLUSIONS. Among the 23,360,953 research records, 5,993,846 records contain at least one structured abstract section as defined above; the remaining 17,287,922 records had English, unstructured abstracts.

## LLM-Based Structuring of Unstructured Abstracts

*Table 2. Adopted prompt templates for the abstract structuring task*

| | |
|---|---|
| **System Prompt** | You are a precise biomedical literature data extraction engine. Your task is to segment a PubMed abstract into a structured JSON format based on NLM categories.<br><br>Strict Constraints:<br>1. Verbatim Extraction: Extract text exactly as written. No paraphrasing or typo correction.<br>2. Header Removal: If the abstract contains explicit headers (e.g., "METHODS:", "STATISTICAL ANALYSIS:"), do NOT include the header labels themselves in the JSON values; extract only the substantive content following them.<br>3. Logical Segmentation: If the abstract is a single block of text (unstructured), use linguistic cues to determine where one NLM category ends and the next begins.<br>4. No Hallucination: If a section is genuinely absent, return null.<br>5. Output: Return a valid JSON object only. |
| **User Prompt** | Extract information from the following PubMed abstract using NLM standard categories.<br><br>Abstract:<br>{abstract}<br><br>Boundary Rules:<br>- BACKGROUND: Context leading up to the gap in knowledge.<br>- OBJECTIVE: Specifically the statement of intent (e.g., "We aimed to...", "This study evaluates..."). If inextricably linked to Background, you may split the text at the transition.<br>- METHODS: Technical steps, participant criteria, and analysis.<br>- RESULTS: The specific data found.<br>- CONCLUSIONS: The final takeaway and interpretation. |

For the 17,287,922 research records with unstructured abstracts, section labels and associated sentences are determined by prompting LLM. By setting verbatim-extraction protocol, Table 2 presents the defined system and user prompts for the abstract structuring task. We evaluate the performance of defined prompts with two model variants: GPT-4.1-nano and GPT-4.1-mini. GPT-4.1-mini is selected as the production model for the full dataset annotation based on superior performance in the technical validation experiment described below. Both GPT models are deployed and served from Microsoft Azure.

## Data Records

We release the corpus, *Structured PubMed*[15], via Zenodo at https://doi.org/10.5281/zenodo.20336717. The corpus consists of two tab-separated values (TSV) files: one containing 5,993,846 records with human-curated structured abstracts, and the other containing 17,287,922 records with LLM-labeled structured abstracts. Each file includes four fields: *pmid*, *structured_abstract*, *type*, and *pubdate*. These metadata, specifically the PubMed ID, publication type, and date, are parsed from the original XML files to ensure unique identification and facilitate downstream metadata retrieval. The *structured_abstract* column contains a JSON object where the keys represent section headers and the values contain the corresponding text. Figure 2 presents an example of the structured abstract.

**Metadata:**

pmid 20392351 | type Journal Article | pubdate 2010-04-16

**Structured abstract:**

BACKGROUND

Tuberculosis (TB) has been associated with poverty, especially in developing countries. Hong Kong has a high incidence of TB where previous reports on the effect of poverty at neighbourhood level have been conflicting.

OBJECTIVE

To examine the spatial distribution of TB and its association with neighbourhood risk factors.

METHODS

A total of 17 294 TB cases notified from 2005 to 2007 were mapped down to the District Council Constituency Area (DCCA) level, and were indirectly standardised by age and sex using 2006 census population data. The standardised TB ratio was correlated with neighbourhood risk factors classified by family, ethnicity, economic and environmental domains

RESULTS

The indirect age- and sex-standardised ratio demonstrated a spatially varied pattern, and was significantly associated with all neighbourhood factors on univariate analysis. Only marital status, place of birth and low household income were independently associated with the standardised TB ratio on multivariate analysis.

CONCLUSIONS

Despite the virtual elimination of absolute poverty by a well-developed social assistance scheme, low household income in the neighbourhood was significantly associated with TB, independently of place of birth, marital status and other risk factors.

*Figure 2 Example of a structured abstract with three metadata fields*

# Technical Validation

## Evaluation Sample

To quantify the accuracy of the GPT-based structuring pipeline, we construct an evaluation framework spanning three distinct PubMed article types: *Research Article*, *Observational Study*, and *Randomized Clinical Trial*. For each article type, we sample 10,000 records (totaling 30,000 records) that contain an author-provided structured abstract with the canonical five-section layout. For each sampled record, the gold-standard section text is known. The concatenation of all five section texts serves as the flattened input abstract fed to the pipeline. This design provides a realistic upper-bound test, where the model receives only the raw, unstructured text that a reader would see and must recover the original section boundaries without any explicit markup or structural cues.

## Metrics

Section-level performance is measured with four standard reference-based metrics computed per section key and averaged across the 10,000 evaluation records:

- ROUGE-1, ROUGE-2, ROUGE-L[16]: unigram overlap, bigram overlap, and longest common subsequence overlap, respectively, between predicted and gold section text.
- BLEU[17]: sentence-level BLEU with effective-order smoothing, which handles short predictions gracefully.

All scores range from 0 to 1 and are computed independently per section key. A macro-averaged mean across the five sections is also reported to summarize model performances. Moreover, we also evaluate a fine-tuned BERT-based sentence classification model (BERT CLS) to benchmark alongside the two GPT models. Specifically, this classification model initializes from PubMedBERT[18] and is fine-tuned using a prompt learning strategy[1] on structured PubMed records.

## Results

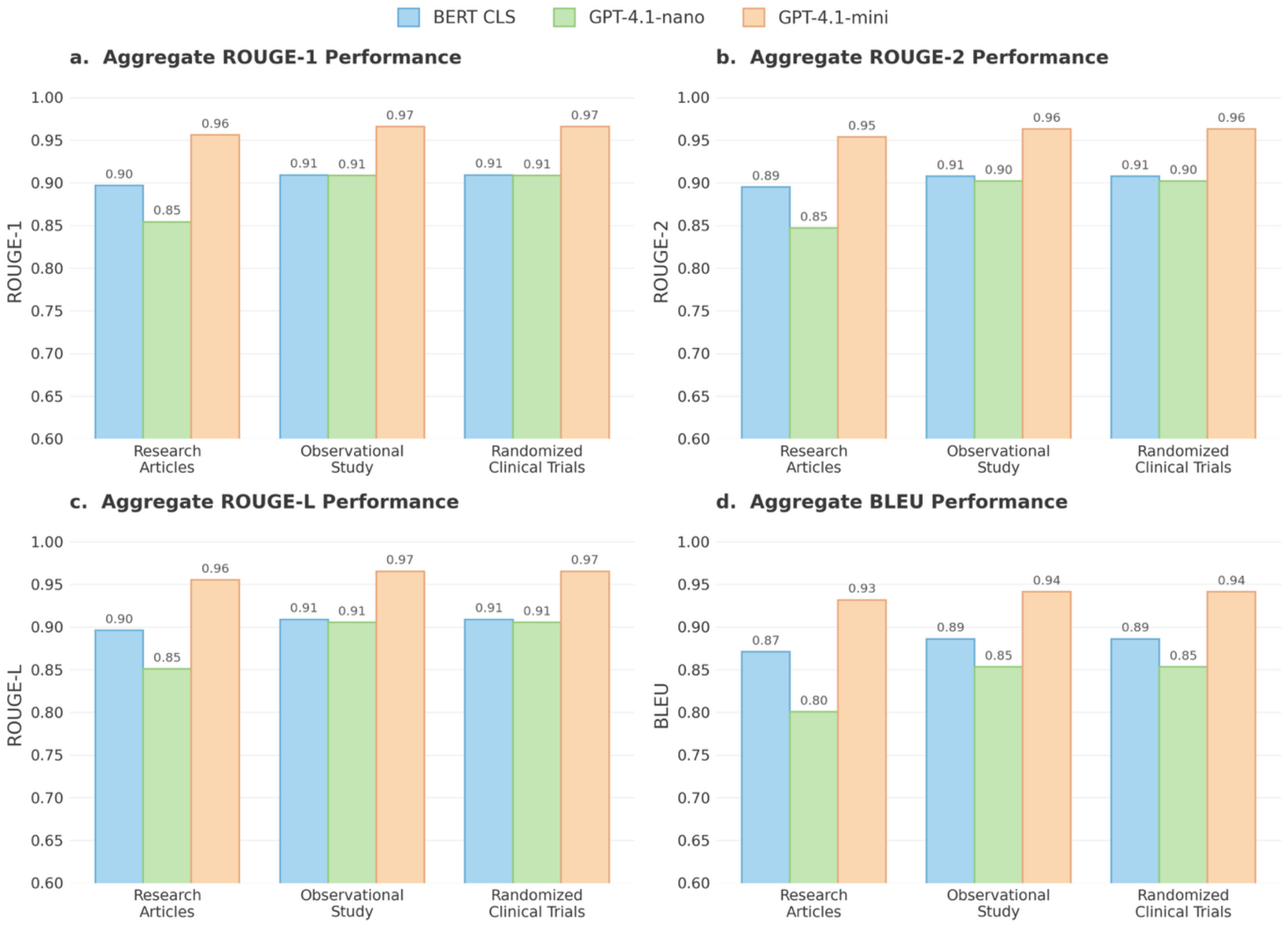

*Figure 3. Aggregate Segmentation Performance Across Article Types. Panels show aggregate results for (a) ROUGE-1, (b) ROUGE-2, (c) ROUGE-L, and (d) sentence-level BLEU scores. Scores denote the mean of performance across the five canonical abstract sections (Background, Objective, Methods, Results, and Conclusions).*

Figure 3 illustrates the aggregate performance of the three models across three different article types. The precise section-by-section breakdown for the *Research Article* samples is detailed in Table 3.

*Table 3. Reference-based evaluation results on 10K research article samples*

| Model | Section | ROUGE-1 | ROUGE-2 | ROUGE-L | BLEU |
|---|---|---|---|---|---|
| BERT-based CLS | BACKGROUND | 0.85 | 0.85 | 0.85 | 0.81 |
| | OBJECTIVE | 0.74 | 0.73 | 0.74 | 0.71 |
| | METHODS | 0.96 | 0.96 | 0.96 | 0.94 |
| | RESULTS | 0.97 | 0.97 | 0.97 | 0.95 |
| | CONCLUSIONS | 0.97 | 0.97 | 0.97 | 0.95 |
| | **Mean** | **0.90** | **0.89** | **0.90** | **0.87** |
| GPT-4.1-nano | BACKGROUND | 0.87 | 0.87 | 0.87 | 0.81 |
| | OBJECTIVE | 0.89 | 0.88 | 0.89 | 0.87 |
| | METHODS | 0.86 | 0.85 | 0.86 | 0.81 |
| | RESULTS | 0.82 | 0.81 | 0.82 | 0.76 |
| | CONCLUSIONS | 0.83 | 0.82 | 0.83 | 0.76 |
| | **Mean** | **0.85** | **0.85** | **0.85** | **0.80** |
| GPT-4.1-mini | BACKGROUND | 0.97 | 0.97 | 0.97 | 0.96 |
| | OBJECTIVE | 0.96 | 0.96 | 0.96 | 0.95 |
| | METHODS | 0.96 | 0.96 | 0.96 | 0.93 |
| | RESULTS | 0.94 | 0.94 | 0.94 | 0.90 |
| | CONCLUSIONS | 0.94 | 0.94 | 0.94 | 0.91 |
| | **Mean** | **0.96** | **0.95** | **0.96** | **0.93** |

## Cross-Type Consistency and Performance

GPT-4.1-mini achieved the highest performance across all evaluation metrics and consistently maintained this superiority across all three document categories. As shown in Figure 3, its aggregate ROUGE-1, ROUGE-2, ROUGE-L, and BLEU scores remain remarkably stable whether processing research articles, observational studies, or randomized clinical trials. For instance, its aggregate BLEU score consistently hovers around the 0.93 to 0.94 range across all article types, demonstrating robust boundary detection that is unaffected by the varying linguistic or structural styles.

## Model Error Profile and Section-Level Granularity

An analysis of the models highlights key differences in how they manage section boundaries across the abstract layouts:

- GPT-4.1-mini maintains exceptionally high, uniform precision across all sections. As shown in Table 3, its scores never dip below 0.90 across any metric or section, demonstrating a consistent capacity for high-level semantic contextualization and precise boundary placement.
- GPT-4.1-nano shows a contrasting error profile: it underperforms in absolute score relative to GPT-4.1-mini. Interestingly, Table 3 reveals that it performs better at isolating introductory elements (BACKGROUND and OBJECTIVE) than BERT-based CLS but experiences a performance drop when delineating the core body and data-heavy sections (METHODS, RESULTS, and CONCLUSIONS).
- BERT-based CLS achieves a high overall average performance (e.g., aggregate ROUGE-1 of 0.91 across most study types in Figure 3), but its performance is highly uneven across specific abstract regions. As highlighted in Table 3, it struggles significantly to isolate the OBJECTIVE section (ROUGE-1 = 0.74), where it routinely confuses background context with the explicit statement of research intent. This confirms that local lexical classification models struggle with boundaries requiring broader semantic context.

Based on its superior absolute performance and exceptional cross-type consistency across all document classes, GPT-4.1-mini was selected for annotating the full corpus of 17,287,922 unstructured research abstracts.

Importantly, the high ROUGE and BLEU scores (particularly for GPT-4.1-mini) reflect the verbatim-extraction design of the prompt: the model is instructed not to paraphrase or summarize, so deviations from the gold standard arise primarily from boundary placement errors rather than lexical substitutions. This property is vital for downstream biomedical informatics applications that require faithful, non-hallucinatory attribution to the original publication text.

# Data Availability

The datasets generated and analyzed in this study are distributed across the following repositories:

- The final 23.3M structured abstract corpus, *Structured PubMed*[15], in TSV format is hosted on the Zenodo at https://doi.org/10.5281/zenodo.20336717. We have added instructions in the README file of our Zenodo repository to demonstrate the use of our corpus.
- The raw source PubMed data can be obtained directly via the official PubMed FTP server[14].
- The 30,000 evaluation samples utilized across the article types (Research Articles, Observational Studies, and Randomized Clinical Trials) for technical validation are available in our GitHub repository: https://github.com/BIDS-Xu-Lab/StructuredPubMed, along with reproduction instructions.

# Code Availability

We have made our codebase publicly available to support reproducibility and further research. The repository, hosted at https://github.com/BIDS-Xu-Lab/StructuredPubMed, includes the script used to parse the raw PubMed FTP XML files, the prompt configurations and the script for extracting structured abstracts via the GPT models, and the evaluation script to reproduce the technical validation results presented in this work.

## Funding

This work received no external funding.